\def\Msun{M$_\odot$}

\def\Teff{T$_{\rm eff}$} 
\def\teff{T$_{\rm eff}$} 
\def\0BMV{(B--V)$_{\rm 0}$} 
\def\BMV{B--V} 
\def\Z{Z} 

\def\V{V}
\def\B{B}

\def\simgt{\lower.5ex\hbox{$\; \buildrel > \over \sim \;$}} 
\def\simlt{\lower.5ex\hbox{$\; \buildrel < \over \sim \;$}}

\documentclass[preprint2]{aastex}
\begin{document} 
%
   \title{Is mass loss along the red giant branch of globular clusters
     sharply peaked? The case of M3}

\author{Vittoria Caloi\altaffilmark{1}   \&  Francesca  D'Antona\altaffilmark{2}   } 
%
%
\affil{\altaffilmark{1}
INAF - IASF Roma, Via Fosso del Cavaliere, I-00133 Roma, Italy;
vittoria.caloi@
iasf-roma.inaf.it
\affil{\altaffilmark{2}INAF - Osservatorio Astronomico  di Roma, via Frascati
33, 00040 Monteporzio, Italy; dantona@mporzio.astro.it}
 }

 
\begin{abstract}
There is a growing evidence that several globular clusters must contain 
multiple stellar generations, differing in helium content. This hypothesis has 
helped to interpret peculiar unexplained features in their horizontal branches. 
In this framework we model   the peaked distribution of the RR Lyr periods in 
M3, that has defied explanation until now. At the same time, we try to 
reproduce the colour distribution of M3 horizontal branch stars. We find that 
only a very small dispersion in mass loss along the red giant branch 
reproduces with good accuracy the observational data.  The enhanced and 
variable helium content among cluster stars is at the origin of the extension 
in colour of the horizontal branch, while the sharply peaked mass loss is 
necessary to reproduce the sharply peaked period distribution of RR Lyr 
variables. The dispersion in mass loss has to be $\le$ 0.003 \Msun, to be 
compared with the usually assumed values of $\sim$ 0.02 \Msun. This requirement 
represents a substantial change in the interpretation of the physical 
mechanisms regulating the evolution of globular cluster stars. 
\end{abstract}

  
\section{Introduction}
Since the early '70s the population distribution along the 
horizontal branch (HB) in globular clusters (GCs) has been interpreted
in terms of a variation of the mass of the hydrogen--rich envelope on
top of the helium core, left over by the red giant evolution. The star
to star difference in the amount of the hydrogen--rich material was
explained as due to the stochastic nature of the mass loss process
\citep{rood1973}. The
main feature to be reproduced was the distribution in {\it number}
of HB members among the red, variable and blue regions. This has been
possible for a large part of the GC system, assuming an average mass
loss along the red giant (RG) branch evolution of $\Delta$M $\sim$
0.22 \Msun, with a $\sigma \sim$ 0.02 \Msun\ \citep[see, f.e.][]{lee1994}.

On the other hand, it has always been clear that a not negligible number
of HBs could not be understood following this assumption. A special case
was that of the exclusively blue HBs (a consistent percentage of the
total GC number), for which an increase in mass loss, in age, in RG
rotation etc., has been invoked, generally with not great success
\citep{fp1993}. But
there are HB distributions that defy any simple explanation: bimodal and
trimodal distributions, extremely long blue tails, hot blue stars in
metal rich clusters are the most impressive examples \citep[e.g.][]{rich1997,
sosin1997,ferraro1998}. 

\begin{figure}[tb]
\centering
   \includegraphics[width=8cm]{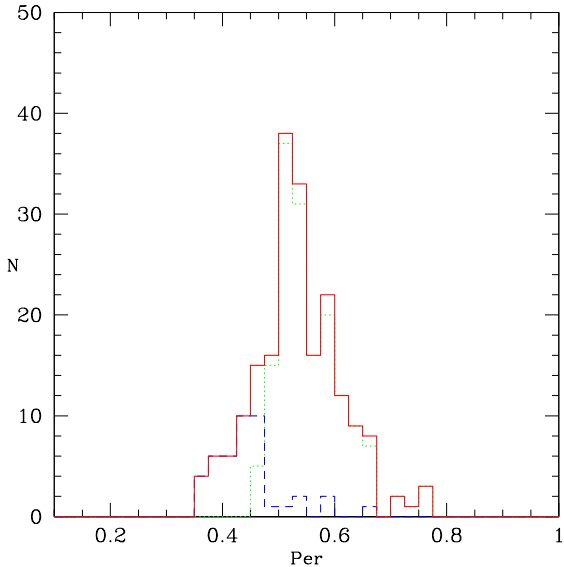}
      \caption{Period histogram (period in days) for M3 RR Lyrs from the data 
      by Corwin \& Carney
     2001, where RRc have been fundamentalized (continuous red line); 
dotted green line: RRabs, dashed blue line: fundamentalized RRcs
}
         \label{fig3}
   \end{figure}
   
\begin{figure}[tb]
   \centering   
   \includegraphics[width=8cm]{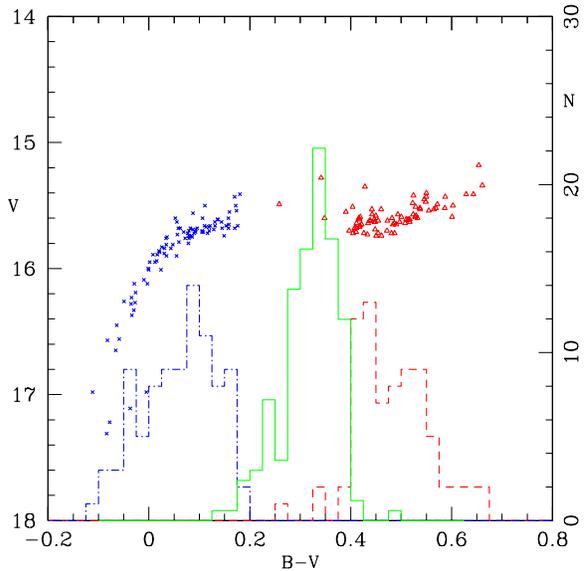}
      \caption{CM diagram of not-variable HB members M3 from the data 
      by Corwin \& Carney
     2001 (not a complete sample, see text), with
     the histogram of their distribution with respect 
    to \BMV. Dashed (red) line: red HB members, dashed-dotted (blue) line:
       blue HB members; continuous (green) line: RR Lyrs histogram}
         \label{fig1}
   \end{figure}

Recently, the presence of peculiar patterns in the chemical abundances
in many (if not all) GCs has provoked a renewed interest in the
hypothesis of helium variations among globular cluster stars 
\citep{norris1981,jb1998,dantona2002}.
The presence of more than one stellar generations, with differing helium
contents, can explain features such as the multiple main sequences in
$\omega$ Cen and NGC 2808 \citep{bedin2004,norris2004,dantona2005,lee2005,piotto2005,
moehler2006,piotto2007}, 
the appearance of blue HB
stars in metal rich GCs (e.g., NGC 6441, Caloi \& D'Antona 2007),
as well as of very blue and very faint HB members that are not
understood in terms of ordinary evolutionary patterns \citep{sweigart2001}.

So a new physical factor, the helium content, has begun to show a
relevance comparable to that of mass loss in shaping HB morphology and
population. At the same time, old unresolved riddles on HB star
distribution, related in particular to some properties of the RR Lyr
variables, have been recalled to attention. The attempts to solve these
problems are imposing unexpected conditions on mass loss properties and
on the role of varying helium content, so that we are possibly facing a
rather substantial change of perspective in the interpretation both of
the HB population and of the mass loss phenomenon during the RG ascent. 
This paper intends to expose this new point of view and
some of its consequences.

\section{M3, the horizontal branch and the RR Lyrae variables} 

The problem of the peaked distribution (see Fig. 1) of the RR Lyr
periods in M3 \citep{ct1981, rc1989} has been revisited by
\cite{catelan2004} and \cite{ccc2005}. Catelan reached the conclusion that
this feature cannot be understood in the framework of canonical HB
evolution, while Castellani et al. found an interesting way to obtain
the observed period distribution with current models. Their solution
requires a strong constraint on mass loss, that has to have a very small
dispersion ($\sigma$ $\sim$ 0.005 \Msun) compared to normally accepted
values ($\sigma$ $\sim$ 0.02 \Msun). This occurrence allows the models
to populate the RR Lyrae region just at the turning point of their
blueward loops, with little dispersion, maximizing their permanence in
the \teff\ interval required to provide the peak in the period
distribution. Anyway, these authors found two main difficulties with
this modelization: i) a number of red HB stars noticeably larger than
observed, and ii) the blue HB stars must be simulated by an {\it ad hoc}
population, with a mean mass, and a $\sigma$\ different from those
required to explain the RR Lyr period distribution.

   \begin{figure}[tb]
   \centering   
   \includegraphics[width=8cm]{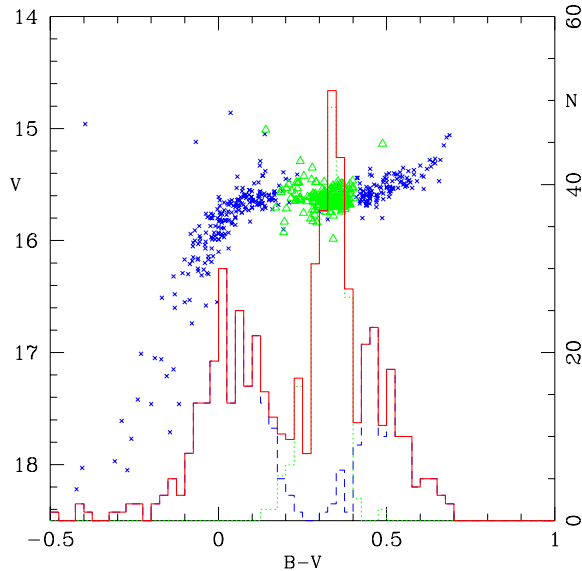}
      \caption{CM diagram of M3 from the data PH94 and CCD96
     for the non variable stars (crosses, blue) plus the RR Lyrs 
     by Corwin \& Carney (triangles, green).
     Also shown is the histogram of their distribution with respect 
    to \BMV (continuous line, red); dashed line (blue): non variable
     stars, dotted (green) line: RR Lyr variables}
         \label{fig2}
   \end{figure}

In this paper we give a model for the period distribution of RR Lyrae
variables in M3, following the suggestion of a very small mass
dispersion for the mass loss along the RG branch.  The extension in
colour of the HB is then reproduced by assuming that the GC contains a
second stellar generation with variable helium content, enhanced with
respect to the first generation. For this cluster, while explaining in
detail the HB morphology and the peaked period distribution of its RR
Lyr, the scenario of multiple star generations provides a strong support
to the rather surprising condition that mass loss must be sharply
peaked.
   
\section{The observational sample}

At present we have periods, average magnitudes and colours for almost
all the RR Lyrae variables in M3 \citep{benko2006, cc2001}. As
guide to the details of M3 HB, we took the photometry by
\cite{cc2001}, who give \V, \BMV\ for about 170 RR Lyr variables (and
periods for 201), together with \B, \V\ photometry for blue and red HB
members. The sample (and the corresponding histograms) shown in
Fig. \ref{fig1} is not complete since it shows only the stars with the
best photometry (see Fig. 4 in Corwin \& Carney), and we have limited the
number of RR Lyrae variables to 100, roughly as expected from HB star
counts in complete samples (see later). The importance of this sample is
that it guarantees a uniform photometry all over the HB, and this is
crucial to estimate the change in number density with colour when
passing from the red to the variable and to the blue regions. From
Fig. \ref{fig1} we see that the number density grows passing from the
red HB to the RR Lyrae region; then there is a dip in the population
between variables and blue HB, followed by another peak at \BMV\ $\sim$
0.0 -- 0.1. The dip in the colour distribution reproduces the paucity of
stars in the RRc colour interval (specially at lower luminosity) and at
the border of the blue region, clearly seen in Fig. 5 in Corwin \&
Carney paper. 
It is important to clarify that this dip is not affected by the well known
uncertainties in the definition of the "mean"colours for RR Lyr variables
\citep[see, f.e.][]{bono1995}. In Fig. 2 we use
magnitude average colours (B-V)$_{\rm mag}$ \citep{preston1961, sandage1990}, 
generally accepted as the best approximation to the colour of
the "static star" with the same average energy output. But any other common
procedure would give the same result for RR Lyrs of type c, owing to their
symmetric light curves \citep{bono1995}, while possible differences up to
0.05 mag may be found for highly asymmetric light curves (as apply to RR Lyrs
of type a). So the relative positions of RRc variables and constant blue HB
stars can be considered as well established.

So any model for the star distribution along the HB must
reproduce these two main features: the sharp peak in colour of the RR
Lyraes (corresponding to the sharp peak in periods), and the dip in the
population at the blue of the variables.
Unfortunately, the \cite{cc2001} HB sample is not complete. We then
examined the complete samples from the CCD photometry by
\cite{ferraro1997} (CCD96) of the external cluster regions, and the
photographic sample PH94 from \cite{buonanno1994}. The colours of the
photographic sample turned out to be very similar, in the red and blue
HB sections, to those in \cite{cc2001}, so that this sample guarantees a
smooth continuity with the colours of the RR Lyrae. To obtain a larger
complete sample, we include also the CCD photometric data by
\cite{ferraro1997}, but we ``corrected back'' this photometry the PH94
photometry by using Equation 7 of the Ferraro et al. paper This
correction is certainly not adequate for the bluest HB stars, which were
not well measured on the photographic plates, but this is not an
important point for our analysis, which is focused on the ``horizontal''
part of the HB.  In Fig. \ref{fig2} we show this complete sample (PH94,
plus CCD96, plus the variables from \cite{cc2001} with the related
population histogram.  The histogram for the variables is normalized to
the number in the PH94 plus CCD96 samples as follows.
The numbers we obtain for the red, variable and blue HB regions are 132,
222 and 217, respectively. This balance among the various populations is
intermediate between what found for inner and outer regions by
\cite{catelan2001}. In fact, they found a substantial difference between
the population distributions for distances from the cluster center r $<$
50'' (28,65,81) and for r $>$ 120'' (27,46,30). Given the uncertatinties
involved, we shall consider our numbers as indicative of a substantial
similarity between the variable and blue populations, while the red
region is noticeably less populated than the variable one. As
already pointed out by \cite{ccc2005}, and as we shall see in the
following, this latter point represents one of the most difficult
constraints with which the simulations have to comply.

Besides, we have to reproduce the RR Lyr period distribution and their
mean period. The sharp peak in the period histogram (Fig. \ref{fig3})
descends from the colour distribution of the variables (Figs. \ref{fig1}
and \ref{fig2}); the mean period $\rm P_{\rm f}$ obtained
fundamentalizing the RRc variables (addition of 0.128 to the logarithm
of their periods, van Albada \& Baker 1973) varies slightly with the
chosen sample, and we assume $\rm P_{\rm f}$ to be comprised between
0.53 and 0.54 d.  The sample by Corwin \& Carney provides $\rm P_{\rm
f}$ = 0.537 d.

\section{The simulations}

\subsection{The models}
We assume for the stars in M3 a heavy element content \Z=0.001 and use
our HB models descending from main sequence structures with $\rm Y_{\rm
MS}$ =0.24, 0.28 and 0.32, as fully described in
\cite{dantonacaloi2004}.  These models are the basis of the syntetic HBs
computed in the following way. We fix a cluster age, and compute the
evolving RG mass according to the relation (1) from
\cite{dantonacaloi2004}, which is a function of both age and Y. We
assume that the stellar content of the cluster is divided into two main
populations: a fraction of stars with cosmological helium (first
generation), for which we assume Y=0.24, and the remaining fraction
having larger and variable helium content (second generation). The
number of the first generation stars, and the number vs. helium
distribution of the second generation stars is changed in order to fit
the whole HB. The HB mass is obtained by assuming that all RG stars lose
an amount of mass $\delta$M, with a gaussian dispersion $\sigma$\ around
a fixed value $\delta$M$_0$. Then the HB mass varies both due to the
mass dispersion $\sigma$\ around the average assumed mass loss, and due
to the dependence of the RG mass on the helium content. As the evolving
mass {\it decreases} with increasing helium content, the stars with
higher helium will populate bluer regions of the HB.
In previous work \citep{dantonacaloi2004} we increased slightly the mass loss
when increasing the helium content, following the idea that the global mass loss
inversely depends on the stellar gravity, and so smaller masses would lose more mass
\citep{lee1994}. Here we assume that the average mass lost is the same for all the
helium contents. In fact, full evolutionary tracks including explicit 
consideration of mass loss, computed according to Reimers' formulation,
showed us that the total mass loss along the RG branch is independent of the
helium content (at each given age): the track location and gravity compensate
in such a way that the mass lost is constant at the level of 1-2$\times 10^{-3}$\Msun,
in most cases.

We assume a parametric mean mass loss along the RG branch with gaussian dispersion
$\sigma$, and extract random both the mass loss and the HB age in
the interval from 10$^6$yr to 10$^8$yr, according to the chosen Y
distribution. We thus locate the luminosity and \Teff\ along
the evolution of the HB mass obtained. These values are transformed into
the observational plane M$_{\rm v}$ vs. \BMV\ using \cite{bessell-castelli-plez1998}.
We identify the variable stars as belonging to a
fixed \Teff\ interval and compute their period according to the
pulsation equation (1) by \cite{dmc2004}. The results are very similar
if we adopt the classic \cite{va1973} relation. The real problem is
given by the choice of the exact boundaries of the RR Lyr strip, that affect
strongly the number and mean period of the RR Lyrae variables.

We assumed as width of the RR Lyr strip $\Delta$log$T_{\rm eff}$ = 0.08,
a value commonly adopted (see, e.g., Castellani et al. 2005), and made a
few experiments varying the choice of the red limit. We found that such
limit {\it had} to be taken at log($T_{\rm eff}$) = 3.80, since for
lower values a much larger than observed number of long period RRab's (P
$>$ 0.6 d) was present. 
Lower limits for the red edge temperature can be
found in the literature (see the already quoted Catelan 2004 and
Castellani et al. 2005). In order to avoid the presence of too many long
period variables, Castellani et al. (2005) make use of decreased values
of their model luminosities. We prefer to keep the theoretical
values for the luminosity and adopt the correspondingly suitable value
for the red edge. 

One has also to consider that the estimate of the limiting temperatures
is by no means a settled question. As discussed, e.g., by
\cite{dmc2004}, the theoretical red edge of the pulsation region gets
hotter by $\sim$ 300 K if the efficiency of superadiabatic convection in the star
external layers is increased by changing the ratio
mixing length to pressure scale height ($l/H_{\rm p}$) from 1.5 to 2.
The comparison with observed periods shows that a
larger convection efficiency should be preferred in several GCs (see
Fig. 14 in Di Criscienzo et al. paper).  At the same time, recent
observational data also suggest hot values for the instability strip red
boundary.  Corwin et al. (1999) found for the colours of the RR Lyr
variables in NGC 5466 (a very metal poor cluster) the limits 0.15 $<$
\0BMV\ $<$ 0.38; similarly, Wehlau et al. (1999) found for NGC 7006, a
GC with a Z close to that of M3, the limits 0.14 $<$ \0BMV\
$<$0.38. Both these colour intervals correspond to \Teff\ intervals very
close to our choice. Besides, Silbermann \& Smith (1995), in their
discussion of the blue and red edges in M15, say that the red edge may
be near 6300 K.

We intend to reproduce the star distribution along the HB in some
detail, not simply to obtain the overall distribution among red (R),
variable (V) and blue (B) regions. To this purpose, the parameters we
can play with are: i) the number of stars with cosmological Y (first
stellar generation); ii) the number of stars with enhanced Y and their
distribution as function of Y; iii) the mass loss and mass loss
dispersion. The total number of stars involved is 571 (see above), to be
distributed among helium contents from the cosmological one (Y = 0.24)
up.
 
\subsection{Results}

On the basis of Fig. \ref{fig1} and Fig. \ref{fig2}, it appeared
reasonable to assume that the red HB and variable regions are populated
mainly by stars with cosmological Y content, while blue HB members
derive from smaller mass RG branch stars with a higher Y
content, starting slightly above the cosmological one. 
evolving RG mass.

Once fixed the age and the heavy element content, for
a given value of Y, the star distribution on the HB is decided by the
amount of the mass loss on the RG branch. After few trials we could
check that the only way to obtain peaked distributions of the variables
in colour and period was to adopt a very well defined value 
for the mass loss. For an age of 11
Gyr and a population with initial Y content = 0.24, a peaked
distribution in (\BMV) (and period, see later) was obtained with a sharp
mass loss value of $\Delta$(M) = 0.203 \Msun, $\sigma$(M) $\leq$
0.003 \Msun. So the mass loss parameter turns out to be the most
delicate and the less arbitrary among the ones mentioned before. We
shall come back to this point.

\begin{figure}[tb]
   \centering   
   \includegraphics[width=8cm]{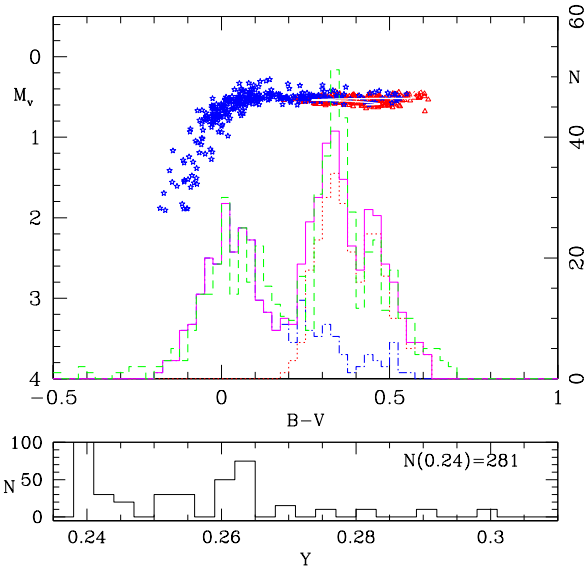}
      \caption{Simulation of the HB distribution (continuous magenta
       line) compared with
	the observed one (dashed green line, see Fig. \ref{fig2}); the
       dotted red line indicates the
	contribution by members with Y=0.24, the dashed-dotted blue line
       the contribution by members with
	enhanced Y, according to the distribution in Y shown in the
	lower panel. In white, the central track of the simulations.}
         \label{fig4}
   \end{figure}

In Fig. \ref{fig4} we show a simulation of the HB distribution obtained
with the quoted parameters, compared
with the observed one. Some points are worth of attention. 
The simulation reproduces the main population peak in the RR Lyr region 
and the smaller one in the red section. It reproduces as well the dip at the
blue of the RR Lyraes and the peak at \BMV\ $\sim$ 0.0. The distribution
in number along the HB gives 159 red members, 211 variable and 201 blue members,
to be compared with the expected 132, 222 and 217 (see Sec. 3.1). Except
for a slight excess of red stars (of the order of two $\sigma$), the
similarity between the observed and the simulated distribution is
remarkable. We stress that this is the first time that a {\it detailed}
simulation has been attempted, since we tried to reproduce not simply
the number of stars in the blue, variable and red HB regions, but the
entire {\it colour histogram} of the HB population.

\begin{figure*}[tb]
   \centering   
   \includegraphics[width=8cm]{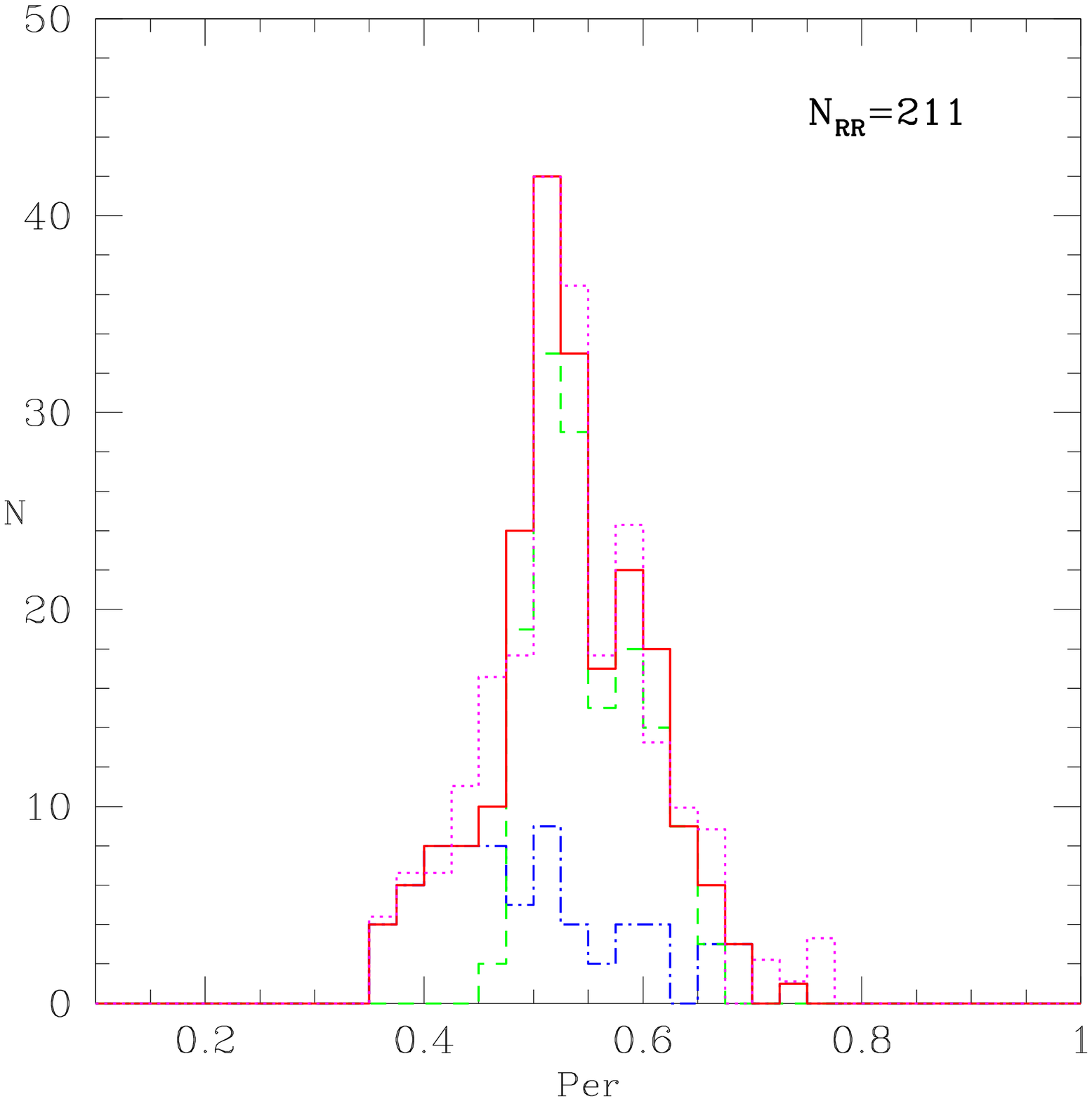}
   \includegraphics[width=8cm]{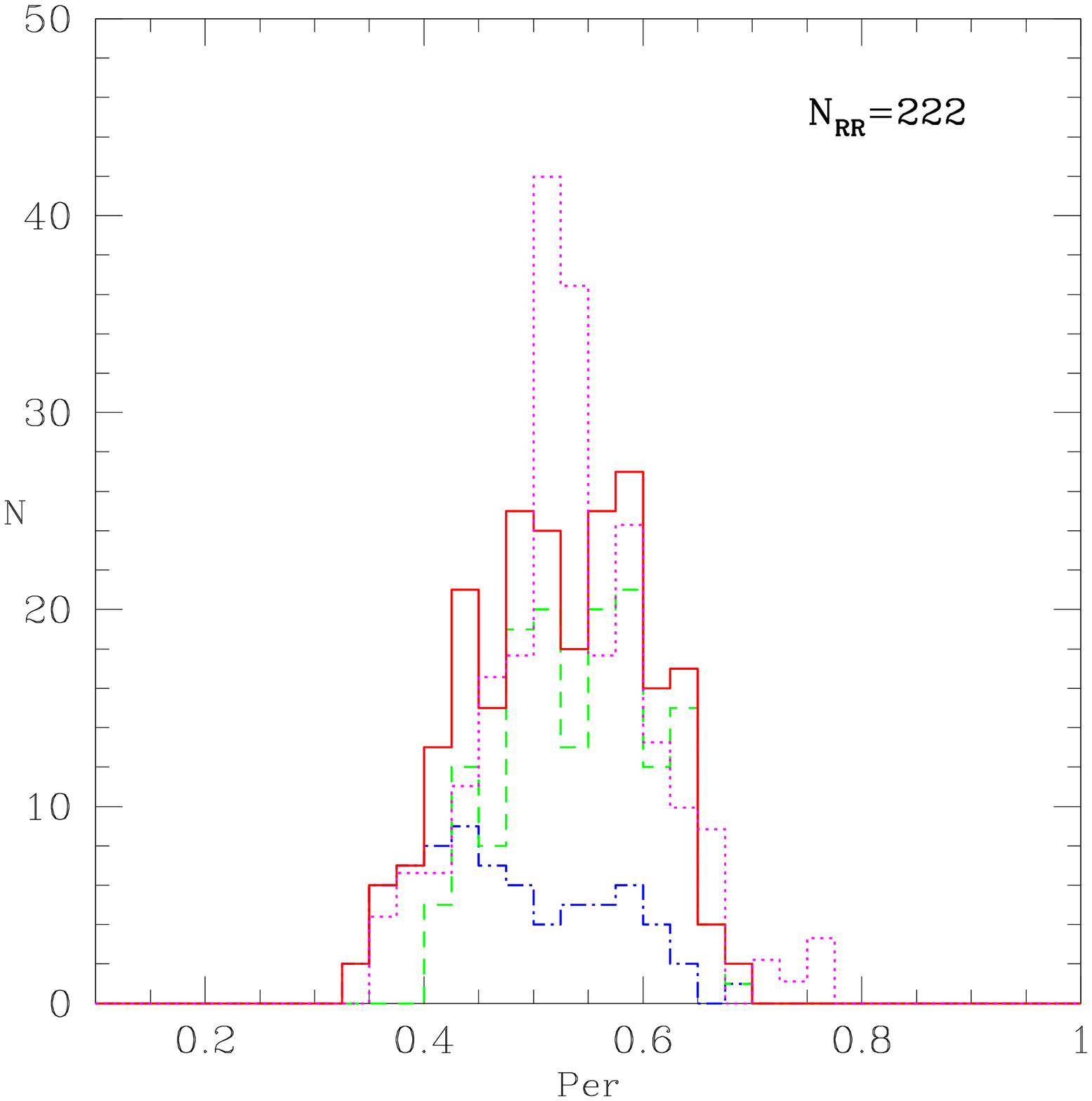}
      \caption{Simulated (continuous red line) and observed (dotted
      magenta line) period
      histograms; the contributions by members with Y=0.24 is indicated
      by the green dashed line, by members with enhanced helium by the blue
      dashed-dotted line. The number of
      ``theoretical'' RR Lyraes is reported. The left panel shows the
simulation for $\sigma=0.0015$\Msun, while the right panel shows the
simulation for $\sigma=0.005$\Msun. All other parameters are the same.}
         \label{fig5}
   \end{figure*}

The left panel of Figure \ref{fig5} shows the period histogram relative
to the same simulation as in Fig. \ref{fig4}, compared with the
observations (dotted line, see Fig. \ref{fig3}).  The observational
histogram of the RR Lyr periods has been obtained with a step of 0.025
d, which allows to appreciate the presence of a secondary peak in the
distribution. We checked this feature comparing the histograms from the
samples by Benk{\H o} et al. (2006) and by Corwin \& Carney (2001), with
an increasing step from 0.01 to 0.04 d. The two distributions appear
very similar, and both show the presence of the
secondary peak for steps $\leq$ 0.03 d, beyond which only the main peak is
evident. Therefore, we consider this feature as well established, and an
intriguing event that our simulation reproduces it. The agreement at
the low and high limits for the period are also quite satisfactory, as
well as the mean period. We think that the simultaneous satisfaction of
so many observational requirements gives strong support to our basic
choices.

Changing $\Delta$(M) from 0.203 \Msun\ to 0.201 or 0.205 \Msun, the
positions of the peaks in Figs. \ref{fig4} and \ref{fig5} appear
slighlty shifted with respect to the observed distributions; in the case
of a lower mass loss, one finds too many red stars ($\sim$ 170). We
cannot exclude these values for what concerns the CM diagram, since the
distribution in \BMV\ cannot be considered completely determined
(photometric imprecision, colour transformations). In any case, a small
change in the absolute value of the mass loss does not affect the main
point, that is, the need for a very small dispersion in mass loss, as
discussed in the following.

 The sharply peaked mass loss allows to have an almost unique value for
  the mass
  evolving through the RR Lyr region, and this is reflected in the peaked
  period distribution. A larger spread in mass loss will give rise to a larger
  spread in the masses evolving through the instability strip, and the sharp
  peak will be smoothed out.

To show how stringent is the assumption of a very small
$\sigma$, the right panel of Figure \ref{fig5} shows the period
distribution for an identical simulation, differing only for the increase in
$\sigma$ to 0.005 \Msun. We see that the period distribution is much
flatter and the peak had disappeared. We applied a Kolmogorov Smirnov
(KS) test to the simulated and observed periods, to derive the
probabilities that they are drawn from the same parent
distributions. The simulation shown in the left panel has a probability
of 89\%, while this is reduced to 5\% for the right panel.  
In order to prove that the situation in Fig. \ref{fig5} is not
peculiar, we applied the KS test to 500 simulations with $\sigma$=0.0015 as well as to 500
ones with $\sigma$=0.005. The simulations have strictly the same inputs, apart from
the dispersion in mass loss. The probability that the observed
and simulated periods are extracted from the same distribution is $>$50\% for 
379 cases (75\%) when $\sigma$=0.0015, but only for 15 cases (3\%) when $\sigma$=0.005.
In the intermediate case of $\sigma$=0.003, we have probability $>$50\% for 245 cases
out of 500. In addition, remember that $\sigma$=0.005 is still a small mass loss dispersion indeed, 
to be compared with $\sigma$=0.02 commonly assumed to reproduce the color extension of HB stars
in M3.

\section{Discussion of the other simulation requirements}

In building up a synthetic HB model with variable helium content, one
might wonder how flexible (arbitrary?) is the choice of the number {\it
vs} helium distribution. Actually, the distribution in number among the
red and variable members on the one side, and the blue HB stars on the
other, depends directly on the assumed number for the first generation
(that is, with cosmological helium) stars. This number has to be lower
than the sum of red and variable stars, because some helium-enriched
stars will always populate these regions, even if originating on the
blue side of the RR Lyr zone. In turn, these stars cannot be too many,
otherwise the RRc region would turn out too populated, with also a
decrease in the mean RR Lyr period. On the other hand, the first
generation cannot be too numerous, otherwise the number of red stars
largely exceeds the expected one.

The details of a helium distribution satisfying these conditions (being
the parent of the simulation discussed in the paper) are reported in the
lower panel of Fig. \ref{fig4}. While minor variations in the
specific numbers would not be relevant, some characteristic features can
be identified: i) the
cluster population is equally divided between original (281) and enriched
(290) helium abundances; ii) the bulk of the helium enriched population is
around Y $\sim$ 0.26; iii) a tail with Y up to $\sim$ 0.30 may be necessary
to explain most of the blue HB members (but we do not wish to put too
much emphasis on this point).

Some stars (about 50) with 0.24 $\simlt$ Y $\simlt$ 0.245 appear in the
helium distribution: they are necessary to avoid too large a dip in
population between RR Lyr variables and blue HB stars. The same position
on the HB would have been attained if we had assumed that these stars
had Y = 0.24 and a slightly larger mass loss. The effect of an increase
in Y from 0.24 to 0.245 is equivalent, for what concerns the resulting
HB star mass, to an increase in the mass lost by an evolving giant, with
Y = 0.24, from 0.203 to 0.210 \Msun. A small asymmetry towards a larger
mass loss would not be unexpected, and would be present in about
15\% of the first stellar generation (50/331).

The amount of blue HB members is given roughly by the number of helium
enhanced stars. As stressed before, there are no definite observational
values for the relative numbers of red, variable and blue stars, and in
any case the number of blue members can be easily accomodated varying
the number of helium enhanced stars. If we exclude the 50 stars with
0.245$\le$ Y $\le$0.24, according to the latter hypothesis, we are left
with $\sim$ 40\% of stars with enhanced helium. This percentage is close
to what we found for other GCs (D'Antona et al. 2006). However, in this
case the helium distribution is peaked at the relatively small value of
Y $\sim$ 0.26, compatible with the mild chemical anomalies observed in
this cluster (Sneden et al. 2004).

\begin{figure}[tb]
   \centering   
   \includegraphics[width=8cm]{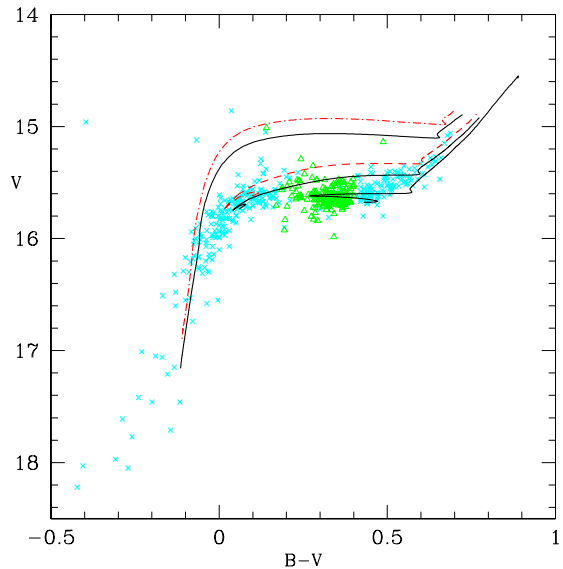}
      \caption{Tracks for standard helium Y=0.24, Z=0.001 and masses M=0.6588, 0.63
        and 0.58 \Msun\ (from right to left, full lines) are superimposed to
        the observational HB data.  The dashed (red) line is the M=0.63 \Msun\ 
       track for Y=0.26, and the dot--dashed (red) line is the M=0.58 \Msun\
       track for Y=0.28. The presence of slightly helium enhanced stars at the
       blue side of the RR Lyrs appears consistent with the data.
       }
         \label{fig6}
   \end{figure}

As most of the HB members in Fig. \ref{fig4} (90\%) has Y $\leq$ 0.265, 
this ensures that the CM diagram of M3 is not
expected to show peculiarities in the position of the main sequence,
the turn--off, etc. In particular, the position of the GB bump will be
very close to that expected for a purely cosmological Y content.
In addition, the HB luminosity of the stars at the blue side of the RR Lyr
region is only mildly affected by the enhanced helium, which at these
locations is in the range 0.25$\simlt$Y$\simlt$0.26. This is shown in Figure
\ref{fig6}, where on the observed HB are superimposed three tracks with
Y=0.24, a track with Y=0.26 and one with Y=0.28.
Maybe a better photometry could discriminate whether actually there is a
helium enhancement at the blue side of the RR Lyrs.

The problem of the red star number has been widely discussed in
Castellani et al. (2005), who found by far too many stars of this
kind. We are less plagued by this problem, but still the number of red
stars in the simulations turns out most of the times larger than
observed. It is not easy to think of a way out of this excess, given the
strict conditions imposed by the RR Lyrae colour and period
distribution. As mentioned before, the contribution to the variables by
stars coming from the blue side cannot be too large because of the dip
in population just at the blue of the RR Lyrae and in the RRc region
(see Fig. 5 in Corwin \& Carney 2001). Besides, too many variables
coming from the blue would lower substantially the observed RR Lyr mean
period. This question will be eaxamined in more detail in the following
Section.

\section{Discussion and conclusions}

\subsection{The global interpretation of the HB of M3}
 
The extraordinary precision in the mass loss required by the simulations
deserves some comments.  In Fig. \ref{fig4} it is shown the track of a
HB star with a mass of 0.659 \Msun, corresponding to the evolving giant
of 11 Gyr (Z=0.001, Y=0.24) after a mass loss of 0.203 \Msun. The two
peaks in the histogram of the star distribution in the red and variable
regions (Fig. \ref{fig4}) have a clear correspondence with the time
development along the evolutionary track. The peak in the red region is
given by the first evolutionary phases along the short redward loop and
the turning toward the blue, while the peak in the RR Lyr region is
the result of the slowing down of the evolution in colour during the
bend of the track at the end of the blueward loop, with the beginning of
the final movement to the red. The requirement of a small dispersion in
mass is due to the necessity of keeping this pattern clear of the
confusion that would arise mixing the evolutions of even a rather small
mass spectrum.

Figures \ref{fig4} and \ref{fig5} show that RRc variables derive mainly
from helium-enhanced stars, at the colour interval around the maximum
blue extension of the 0.659 \Msun\ track. Here one finds also the dip in
the star distribution at the RR Lyr blue limit, set by the colour
limit of HB members with original helium content, and by the small
difference in helium content among the first star generation and the
succeeding ones. The dispersion along the blue HB, given the reduced
dispersion {\it in mass}, is due to the dispersion {\it in helium}, as
shown in the lower panel of Fig. \ref{fig4}. 

In fact, there are two main differences between our analysis and the
analysis by \cite{ccc2005}: i) they assume for the mass distribution
which provides the RR Lyr variables a $\sigma=0.005$ \Msun, while we
have shown that our best matches with the period distribution are
obtained by a $\sigma$\ much smaller -- almost a unique mass is evolving
in the RR Lyr region; ii) they have to fill up the blue side of the HB
with an {\it ad hoc} population, while we assume the model of a second
stellar generation with varying helium content, as we have already done
to interpret other complex HB distributions, such as in NGC 2808
\citep{dantonacaloi2004} and in NGC 6441 \citep{cd2007}. In this
framework, our interpretation for the strong constraint on the mass loss
rate along the RG is much more cogent. 

\subsection{A sharp value of mass loss also in other GCs?}

The population distribution along the HB and the period distribution of
the RR Lyr variables have been obtained assuming the presence of (at
least) two populations with differing helium contents and, most
important, an extremely narrow dispersion in the mass loss. This latter
condition may be considered as necessarily requiring the presence of
the first condition, since the blue HB population cannot be
understood in terms of ``one mass'' evolution, as it happens for the red
and variable members.

So the multiple star generations with varying Y hypothesis appears
capable of coping with the rather strong condition of a very sharp
amount of mass loss during RG evolution in M3
population. The variations in Y takes the place of the variations in
mass, to explain the HB star distribution. We have seen already in
other clusters similar situations. For example, in NGC 2808 (D'Antona \&
Caloi 2004) the red clump required a relatively small mass dispersion
($\sigma \simeq$ 0.015), and the long blue HB region was obtained with a
distribution in Y. But the situation in M3 appears much more stringent
in the requirement on mass loss, basically due to the RR Lyr properties
that cannot be satisfied with a not--peaked mass distribution.

How special is the case of M3? The point has been discussed by Catelan
(2004) and Castellani et al. (2005), who incline to consider M3 as
perhaps a ``pathological'' case. They observe that M5 and M62, both rich
in RR Lyr variables, show a less pronounced peak in the period
distribution (see also Castellani et al. 2003, Fig. 1). While this is perhaps
true, at least for M5, one finds in the quoted Fig. 1 (Castellani et
al. 2003),
many other Oosterhoff I clusters with about 70 RR variables (as in M62,
according to Clement et al. 2001) with strongly peaked period
distributions (NGC 6715, 6934, 7006). The problem with these clusters
appears similar to that found in M3. Let us quote also the case of
Palomar 3 (Catelan et al. 2001), that requires a very small mass
dispersion (consistent with zero, according to the authors) to describe
the HB distribution.

We think therefore that the problem is with us to stay. If the small
dispersion in mass loss will be confirmed, we will face a rather
impressive change of perspective in the interpretation of HB
distributions. The role of the first phases of the dynamical and
chemical evolution of GCs will become more and more crucial. In turn,
these phases are dominated by the behaviour of the initial mass
function, of the chemical processes in asymptotic giant branch stars and
on their mass loss efficiency. So, on the one side we think that the
hypothesis of multiple stellar generations begins to find a certain
amount of observational support, and on the other we are well aware of
the long way to reach an understanding of how such a phenomenon took place.

\acknowledgements
We thank Drs. C. Corsi, T.M. Corwin and F. Ferraro  for kindly sending
us their data on variable and not variable stars in M3 horizontal
branch, and Dr. M. Castellani for useful discussion. 


\end{document}